\documentclass{amsproc}
\usepackage{graphicx}

\theoremstyle{definition}

\theoremstyle{remark}

\numberwithin{equation}{section}

\begin{document}

\title{Resolving Gravitational Singularities}

\author{Finn Larsen}
\address{Department of Physics, University of Michigan, Ann Arbor MI-48109, USA.}
\curraddr{Theory Division, CERN, CH-1211, Geneva 23, Switzerland.}
\email{larsenf@umich.edu}
\thanks{The author was supported in part by DoE grant \# DE-FG02-95ER40899.}


\date{August, 2008.}


\begin{abstract}
We review recent progress on the resolution of gravitational singularities in string theory. 
The main example is the fundamental string in five dimensions which is singular in the 
standard supergravity description but regular after taking into account higher derivative 
corrections determined by anomalies and supersymmetry. The application of the AdS/CFT
correspondence to geometries that require higher derivatives for regularity poses
interesting challenges. 
\end{abstract}

\maketitle

.

\section{Introduction}
In the last several years there has been significant progress on understanding higher 
derivative corrections to black holes in string 
theory \cite{Behrndt:1998eq,Lopes Cardoso:1998wt,Lopes Cardoso:1999ur,LopesCardoso:2000qm,Ooguri:2004zv,Verlinde:2004ck}. An important aspect of 
this development is that some interesting examples have been identified
where gravitational singularities are resolved by higher 
derivative corrections to the action 
\cite{Dabholkar:2004yr,Dabholkar:2004dq,Hubeny:2004ji,Sen:2004dp,Dabholkar:2005by,Kraus:2005vz,Kraus:2005zm}. The prototypical example is a regular black 
hole that would have vanishing horizon area were it not for the higher 
derivative corrections in the Lagrangian. The resolution of gravitational singularities
is an important problem in many different contexts so it is worthwhile discussing this 
aspect of higher derivative corrections separately from perturbative corrections to solutions
that are already regular in the leading approximation. That is the purpose of this lecture. 

The work discussed here is collected from a series of articles by Castro, Davis, 
Kraus, and the current author \cite{Castro:2007sd,Castro:2007hc,Castro:2007ci}. These articles
develop the subject in the context of five dimensional supergravity corrected by the mixed gauge-gravitational Chern-Simons 
term:
\begin{equation}
{\mathcal L}_{\rm CS} = {1\over 24\cdot 16\pi^2} ~c_{2I}A^I\wedge {\rm Tr} ~R\wedge R~,
\label{eq:ano}
\end{equation}
and terms related to this by supersymmetry \cite{Hanaki:2006pj}. In this lecture we sacrifice 
generality and technical detail in order to keep the text brief and, hopefully, accessible
also for interested non-experts. A much more comprehensive review appeared recently
in \cite{Castro:2008ne}.

Before considering any specifics, it is useful to recall some of the general challenges
intrinsic to any attempt at resolving singularities using higher derivative terms:
\begin{itemize}
\item
{\it Exact symmetries} of the theory must be preserved by the higher derivative corrections.
In our context, the Chern-Simons term (\ref{eq:ano}) is required for anomaly cancellation. 
Furthermore, we go through much effort to preserve supersymmetry. 
\item
{\it Effective field theory} demands that {\it all} terms of a given order should be taken into account. 
We address this by appealing to the uniqueness of the supersymmetric completion of the 
mixed gauge-gravitational Chern-Simons term (\ref{eq:ano}).
\item
{\it Field redefinition ambiguities}, such as $g_{\mu\nu} \to g_{\mu\nu} + AR_{\mu\nu}$
(and generalizations involving matter fields), mix terms of different derivative order
so that the singularity 
could depend on the choice of coordinates in field space.
Our construction employs an off-shell formalism that implements supersymmetry at each 
order in the derivative approximation independently. This precludes most field redefinitions 
and defines the metric unambiguously. 
\item
{\it Singularities} makes any systematic expansion inherently problematic. In 
order to cancel leading order singularities, the higher order terms must be comparable 
to the leading order terms. Then terms that are formally of even higher order 
might be important as well,
{\it i.e.} there is no systematic expansion parameter. We will not be able to address 
this point completely but we will make it very explicit and concrete in due course. 
\end{itemize}
We now turn to the case of small five-dimensional black strings in $N=2$ supergravity, the 
example that is the focus of this talk. 

\section{Singularities or no singularities?}
Consider M-theory on $CY_3\times R^{4,1}$ for small $CY_3$ volume. 
The theory is 
effectively five dimensional and in the supergravity approximation it reduces to 
$N=2$ supergravity in $D=5$ coupled to a number of vector multiplets. The Lagrangian is
\begin{equation}
{\mathcal L}_0 = R + G_{IJ}\nabla M^I \nabla M^J + {1\over 2} G_{IJ} F^I_{ab} F^{Jab}
- {1\over 24e} c_{IJK} A^I_a F^J_{bc} F^K_{de} \epsilon^{abcde}~,
\label{ba}
\end{equation}
in a self-explanatory notation (the theory is introduced in more detail in {\it e.g.} \cite{Larsen:2006xm}). 

We seek string solutions to this theory and so consider the {\it ansatz}:
\[ 
ds^2 = e^{2U_1(r)} \left(dt^2  -  dy^2\right)  - e^{-4U_2(r)} (dr^2 + r^2 d\Omega^2_2)~,
\]
for the geometry. The strings are supported by magnetic charges $p^I$ with respect to the vector 
fields $A^I$, {\it i.e.} the two forms $F^I$ 
\begin{equation}
F^I = - {1\over 2} p^I  e^{\hat\theta}\wedge e^{\hat\phi}~,
\label{eq:fi}
\end{equation}
give rise to magnetic flux through the transverse two-sphere. 

The warp factors $U_{1,2}(r)$ are going to diverge at $r=0$ so there will at 
least be a coordinate singularity near the string. However, the singularity may just
be a coordinate artifact, the actual geometry could be regular. This happens 
when $e^{-6U_1}=e^{-6U_2} \sim r^{-3}$ as $r\to 0$. Moreover, the resulting
regular solutions have $AdS_3\times S^2$ near string geometry. 

The actual behavior of the string solution depends on the charge configuration. 
An important special case is when the matter supporting the string solution is the two-form 
$B_{\mu\nu}$ (represented by the one-form gauge-field that it dualizes to in five dimensions) 
and the dilaton $\Phi$, as in 
perturbative string theory. In this case the solution to the supergravity equations of motion 
has $e^{-6U_1}=e^{-6U_2} \sim r^{-1}$ as $r\to 0$. This solution is therefore singular 
in the supergravity approximation. We want to establish that higher derivative corrections 
to the theory can modify the warp factors near the string so that in fact 
$e^{-6U_1}=e^{-6U_2} \sim r^{-3}$ as $r\to 0$, leaving the full geometry regular. 

For the class of string solutions considered here there is a simple and
general criterion determining whether
the geometry has a singularity. If the charge configuration is such that 
$c_{IJK} p^I p^J p^K$ is nonvanishing, then there are 
magnetic string solutions to standard (two-derivative) supergravity with the
regular $AdS_3\times S^2$ near string geometry with scale set by $c_{IJK} p^I p^J p^K$.
However, if the charges are such that $c_{IJK} p^I p^J p^K=0$, then the classical 
geometry is singular. 

The string solutions in five dimensions are interpreted in M-theory as  
$M5$-branes wrapped on four-cycles $P$ in $CY_3$. The magnetic 
charge $p^I$ is the wrapping number around the basis four-cycle $P_I$. 
An $M5$-brane wrapping a general four-cycle $P$, one not coinciding with 
any basis-cycle $P_I$, carries several (or many) of the charges $p^I$. 
The important combination $c_{IJK} p^I p^J p^K$ can be interpreted as
the self-intersection number of the four-cycle $P$ underlying the string solution, 
because the coupling constants $c_{IJK}$ 
in the action (\ref{ba}) are interpreted microscopically as the intersection 
numbers of the four-cycles $P_I$.

We will be particularly interested in the special case where the $CY_3$ is the product manifold
$K3\times T^2$ and the $M5$-brane wraps the four-cycle $P=K3$. This solitonic string is 
important because it is the type IIA dual of the heterotic string. Since the four-cycle
underlying the dual heterotic string corresponds to one of the basis cycles it has 
vanishing self-intersection number $c_{IJK} p^I p^J p^K=0$. The criterion using
the self-intersection number $c_{IJK} p^I p^J p^K=0$ thus reproduces the result
that heterotic strings have singular near string geometry in the supergravity approximation.

\section{Indirect Resolution of Singularities: Anomalies}
The magnetic string solutions we consider are subject to powerful symmetry 
principles. These principles suggest when we should expect that
gravitational singularities are resolved. 

\subsection{AdS/CFT correspondence}
We must first recall the significance of the $AdS_3\times S^2$ 
geometry. The global isometry group of $AdS_3$ is:
\[
SO(2,2)\simeq SL(2)\times SL(2)~.
\]
This isometry induces an obvious global $SL(2)\times SL(2)$ symmetry acting on the 
asymptotic boundary of $AdS_3$. The important point is that there is much more
symmetry\cite{Brown:1986nw}: bulk diffeomorphisms combine with the global $SL(2)\times SL(2)$
and enhance the full boundary symmetry to ${\mathrm Vir}\times {\mathrm Vir}$. 
Similarly, the $SO(3)\simeq SU(2)$ isometry group of the $S^2$ combines 
with bulk diffeomorphisms and form the affine current algebra $\widehat{SU(2)}$ 
acting on the left-movers (the supersymmetric side) in the two-dimensional boundary 
CFT. 

The bulk spacetime supersymmetry complements the bosonic isometries so that there
is in fact a superisometry-group. Combining this with bulk diffeomorphisms one finds
a superconformal algebra on the boundary. In the case of $N=2$ supersymmetry in bulk 
the correct superisometry is $SU(1,1|2)$ and the boundary theory becomes a
$(4,0)$ superconformal CFT. 

Explicit computation from the standard (two-derivative) supergravity action (\ref{ba})
determines the spacetime central charges of the Virasoro algebras\cite{Brown:1986nw}:
\begin{equation}
c_L = c_R = {3\ell_A\over 2G_3}~,
\label{eq:dg}
\end{equation}
where $\ell_A$ is the length scale of the $AdS_3$ space. The central charge is a 
measure of the 
number of degrees of freedom in the two-dimensional
boundary CFT. In bulk it is essentially the size of $AdS_3$, suggesting that
we can use central charge as a convenient proxy for the scale of the geometry. 

The affine current algebra $\widehat{SU(2)}$ is similarly found to have level $k= c_L/6$,
consistent with the $(4,0)$ superalgebra. Geometrically, this relates the scale
of the $S^2$ to that of AdS$_3$. We will later see that the precise relation is
$\ell_A=2\ell_S$. 

Expressing $\ell_A$ in terms of the magnetic charges $p^I$ of the string, it can be 
shown that the supergravity central charge (\ref{eq:dg}) can be written as:
\begin{equation}
c_L = c_R = c_{IJK} p^I p^J p^K~.
\label{eq:dh}
\end{equation}
In other words, in the supergravity approximation the scale $\ell_A$ is essentially 
the intersection number $c_{IJK} p^I p^J p^K$, which in turn is the central
charge.

We previously noted that the vanishing (or not) of the intersection number 
$c_{IJK} p^I p^J p^K$ gives a criterion for whether the solution is singular (or regular). We 
now see that how this fits with the relation between central charges and the size of 
the AdS$_3$ (and of $S^2$). For charge configurations such that the 
supergravity central 
charge (\ref{eq:dh}) vanishes, the corresponding scale vanishes and the 
geometry is singular. 

We can now explain why we should expect that singularities are resolved:
even when the classical central charge (computed in the supergravity approximation)
vanishes, it is reasonable to expect that the exact central charge (including
corrections) does not vanish. This motivates the expectation that the geometry 
should be regular when higher derivative corrections are taken into account. 

\subsection{Local Anomalies}
The most robust higher derivative corrections are those related to anomaly cancellation.
The standard Green-Schwarz terms (or alternatively terms inferred from M5-brane anomaly 
cancellation) give rise in five dimensions to the mixed gauge-gravitational Chern-Simons term
\begin{eqnarray}
{\mathcal L}_{\rm CS} &=& - {c_{2I}\over 96e} A^I \wedge {\rm Tr} R\wedge R\cr
&=&  {c_{2I}\over 96e} F^I \wedge \omega_3+{\rm tot.der.}
\label{eq:da}
\end{eqnarray}
In the first line the interaction was written in a form that violates gauge symmetry, 
albeit only by a total derivative. 
The second lines introduces the Chern-Simons three-form $\omega_3$ (through 
${\rm Tr} R\wedge R=d\omega_3$) to make the term manifestly gauge invariant
but then diffeomorphism symmetry is violated, albeit again just by a total derivative.

The AdS/CFT correspondence relates the diffeomorphism violations in bulk to 
corresponding anomalies in the boundary theory \cite{Kraus:2005vz,Kraus:2005zm}. 
The relation determines the boundary central charges as:
\begin{equation}
c_L =  c_{IJK} p^I p^J p^K  + {1\over 2} c_{2I}p^I~,~~~~c_R =  c_{IJK} p^I p^J p^K  + c_{2I}p^I~,
\label{eq:cd}
\end{equation}
in accordance with other arguments \cite{Maldacena:1997de,Harvey:1998bx}. 
Both central charges are determined because diffeomorphism symmetry in AdS$_3$
and in $S^2$ (which is R-symmetry in the dual theory) give two pieces of information. 
The expressions (\ref{eq:cd}) for the central charges are exact because the local 
symmetries that they enforce are exact. Even though supersymmetry plays 
an important role in the reasoning (relating $R$-charge and central charge on the supersymmetric side)
we ultimately determine $c_L$ (the supersymmetric side) and $c_R$ 
with equal precision. 

\subsection{Resolution of singularities}
The exact central charges (\ref{eq:cd}) suggests the resolution of gravitational singularities. 
For this we consider charge vectors such that the intersection number 
$c_{IJK}p^I p^J p^K=0$, corresponding to singular geometry in the leading order
description. In this case the central charges (\ref{eq:cd}) are linear in 
the magnetic charges
\begin{equation}
c_L =   {1\over 2} c_{2I}p^I~,~~~~c_R =   c_{2I}p^I~,
\label{eq:ce}
\end{equation}
and in general they do not vanish. To the extent that central charges can be taken as 
a measure of the scales of the near horizon $AdS_3\times S^2$, we see that higher 
derivative corrections have replaced a singularity (a space with vanishing size)
with a regular geometry (with nonvanishing scale). 
Thus a gravitational singularity has been resolved.

It is of special interest to consider the case of the dual heterotic string. Recall
this is the case where the compact Calabi-Yau three fold is $CY_3=K3\times T^2$
and the $M5$-brane simply wraps the four-cycle $P=K3$. Since $c_2(K3)=24$ 
the central charges (\ref{eq:ce}) become
\[
c_L = 12p~~~,~c_R=24p~. 
\]
These are the correct values for $p$ heterotic strings in a physical gauge. (The 
left movers are $8$ bosons
and $8$ fermions with central charge $c_L = 8\cdot 1 + 8\cdot {1\over 2} = 12$, 
the right movers are $24$ bosons with central charge $c_R= 24\cdot 1=24$).
This gives confidence that both the central charges (\ref{eq:ce}) have been 
correctly determined. It also suggests that the 
gravitational singularity has been resolved in the specific case of the heterotic string. 

 

%
%

\section{Explicit Singularity Resolution}
So far we have used symmetries to argue that certain gravitational singularities are 
resolved. However, there are several ways 
the argument could fail. For example, we {\it assume} a $AdS_3\times S^2$ 
near string geometry and then anomalies and symmetries determine the central 
charges. This reasoning would fail if the near string geometry was not 
AdS$_3\times S^2$. Another point is that we use central charges as a proxy for 
the scale of the geometry and this relation is not precise when higher derivative
corrections are taken into account. In order to understand these and other issues 
better we would like to explicitly construct asymptotically flat solutions with the 
anticipated properties. This is what we turn to next. 

\subsection{Supersymmetry}
As we have already mentioned repeatedly, the essential ingredient is the mixed 
gauge-gravitational Chern-Simons term (\ref{eq:da}). Since the indirect arguments 
rely in part on supersymmetry, all terms related to the 
Chern-Simons term by supersymmetry must also be taken into account. 

The direct way to supersymmetrize (\ref{eq:da}) would be to carry out the Noether procedure:
act by the standard supersymmetry transformations and identify other terms in the Lagrangian. 
Then modify the supersymmetry transformations as needed and improve
the supersymmetrization iteratively. 
The problem is that this procedure generates terms of arbitrarily high order so that there is
no closed form of the Lagrangian.

This challenge can be addressed by using off-shell supermultiplets identified 
using the superconformal formalism. In the gauge fixed version of the 
superconformal formalism that we employ, the only complication relative
to the on-shell formalism is that the standard
physical fields must be augmented by auxiliary fields. In particular, the Weyl multiplet
({\it i.e.} gravity) has an auxiliary scalar $D$, auxiliary two-tensor $v_{ab}$, 
and also an auxiliary fermion $\chi$. 

The off-shell action is invariant under the supersymmetry transformations:
\begin{equation}
\begin{split}
\delta\psi_\mu & = \left( {\mathcal D}_\mu + {1\over 2} v^{ab}\gamma_{\mu ab} - {1\over 3} \gamma_\mu
\gamma\cdot v\right)\epsilon~, \cr
\delta\Omega^I & = \left( - {1\over 4}\gamma\cdot F^I - {1\over 2} \gamma^a \partial_a M^I
- {1\over 3} M^I \gamma\cdot v\right) \epsilon~, \cr
\delta\chi & = \left( D - 2\gamma^c \gamma^{ab}{\mathcal D}_a v_{bc} - 2\gamma^a\epsilon_{abcde}
v^{bc} v^{de} + {4\over 3} (\gamma\cdot v)^2\right) \epsilon~.
\label{eq:BPS}
\end{split}
\end{equation}
It is important to emphasize that these transformations 
are symmetries of each order in the action by itself. Results we find by analyzing 
supersymmetry will therefore apply to the leading order solution, and also to the 
solution corrected by four derivative terms. Indeed, they must  apply to the exact 
solution which take into account corrections to all orders. 

The supersymmetry variations (\ref{eq:BPS}) take a form similar to the standard
on-shell supersymmetries, except for the unfamiliar appearance of the auxilary fields.
We exploit these variations in the familiar way: demanding that 
the supersymmetry transformations vanish, when evaluated on purely bosonic backgrounds,
yields linear differential equations satisfied by BPS-solutions.

As advertized previously, we assume as an {\it ansatz} that the metric takes the string form:
\[
ds^2  = e^{2U_1} ( dt^2 - dy^2) - e^{-4U_2} ( dr^2 + r^2 d\Omega^2_2)~.
\]
Now the BPS conditions (\ref{eq:BPS}) impose $U_1=U_2$ (so we can omit the index $1,2$ on $U$) 
and also determine the auxiliary fields in terms of the metric 
function $U$: 
\begin{equation}
\begin{split}
v & = {3\over 4}e^{2U} \partial_r U e^{\hat\theta}\wedge e^{\hat\phi}~,\cr
D & = 6 e^{4U} \nabla^2 U~,
\end{split}
\end{equation}
where $\nabla^2$ is the Laplacian on the three-dimensional space transverse to the string. 
Finally, the BPS condition relates the magnetic fields and the scalar fields:
\begin{equation}
F^I = {1\over 2} \partial_r\left( M^I e^{-2U} \right) e^{4U}  e^{\hat\theta}\wedge e^{\hat\phi}~.
\label{eq:FI}
\end{equation}
This is the standard attractor flow which represents field strengths as gradient
flows, with the scalar fields essentially acting as potentials. 

\subsection{The Bianchi identity}
 
The scalar fields $M^I$ and the metric function $U$ are not determined by supersymmetry 
alone. Generally, we must at this point appeal to the equations of motion which depend on 
the action. However, the case of 5D strings is special because it is magnetic fields  
\[
F^I = - {1\over 2} p^I  e^{\hat\theta}\wedge e^{\hat\phi}~,
\]
that support the string solution. These fields are topological and so they must be exact. 
However, the field strengths (\ref{eq:FI}) determined from supersymmetry do not automatically
take the correct form; they may even fail to satisfy the Bianchi identity. Imposing the
Bianchi identity we find the harmonic equation
\[
\nabla^2 \left( e^{-2U} M^I\right) =0~. 
\]
In the spherically symmetric case assumed here
the solution is 
\[
M^I e^{-2U} = H^I = M^I_\infty + {p^I\over 2r}~,
\]
where $M^I_\infty$ are integration constants. 
These are the harmonic functions which underlie many familiar two derivative solutions.
Since we have not yet specified the action, these are in fact {\it precisely} the
standard harmonic functions, with normalizations and other details unchanged from the
two derivative case. 

\subsection{The equations of motion}
Up to this point we used supersymmetry and the Bianchi identity to determine the magnetic 
string solution completely, except for the metric factor $U(r)$. It is this function that 
depends on the detailed action. 
In order to get some familiarity with the off-shell formalism we first introduce the
two-derivative action
\begin{equation}
\begin{split}
{\mathcal L}_0 & = -{1\over 2} D + {3\over 4}{\mathcal R} + v^2
+{\mathcal N}\left( {1\over 2}D + {1\over 4} {\mathcal R} + 3v^2\right)
+ 2{\mathcal N}_I v^{ab} F_{ab}^I\cr
& + {\mathcal N}_{IJ} \left( {1\over 4} F_{ab}^I F^{Jab} + {1\over 2}\nabla M^I \nabla M^J\right)
+ {1\over 24e} c_{IJK} A^I_a F^J_{bc} F^K_{de} \epsilon^{abcde}~. 
\label{eq:lag0}
\end{split}
\end{equation}
This action appears much more complicated than the standard supergravity action (\ref{ba}). 
However, the auxiliary fields $D$ and $v_{ab}$ enter algebraically so they can be integrated
out exactly, by imposing their equation of motion. The equations of motion
for $D$ gives
\begin{equation}
{\mathcal N} = {1\over 6} c_{IJK} M^I M^J M^K = 1~. 
\label{eq:specgeom}
\end{equation}
This is the familiar special geometry constraint. It ensures that the number of independent
physical scalar fields is one less than the number of vector fields, as must be the case because one of
the vector fields is the graviphoton which has no scalar superpartner. 
Simplifying the Lagrangian (\ref{eq:lag0}) using the special geometry constraint 
(\ref{eq:specgeom}) and the
equation of motion for $v_{ab}$ we recover the familiar on-shell $N=2$ Lagrangian (\ref{ba}) 
with the correct expression 
\[
G_{IJ} = {1\over 2} \left(  {\mathcal N}_I {\mathcal N}_J- {\mathcal N}_{IJ} \right)~,
\]
for the metric on moduli space. 

We are now ready to introduce the four derivative action in all its glory \cite{Hanaki:2006pj}
\begin{equation}
\begin{split}
{\mathcal L}_1& = {c_{2I} \over 24}
\left( {1\over 16} \epsilon_{abcde} A^{Ia} R^{bcfg}R^{de}_{~~fg} +
{1 \over 8}M^I C^{abcd}C_{abcd} +{1 \over 12}M^I D^2 +{1 \over
6}F^{Iab}v_{ab}D \right. \cr
& ~~+{1 \over 3}M^I C_{abcd}
v^{ab}v^{cd}+{1\over 2}F^{Iab} C_{abcd} v^{cd}
+{8\over 3}M^I v_{ab} {\mathcal D}^b {\mathcal D}_c v^{ac} -\frac{16}{9} M^I v^{ac}v_{cb}R_a^{~b} \cr
&~~-\frac{2}{9} M^I v^2 R +{4\over 3} 
M^I {{\mathcal D}}^a v^{bc} {{\mathcal D}}_a v_{bc} + {4\over
3} M^I {{\mathcal D}}^a v^{bc} {{\mathcal D}}_b v_{ca} -{2\over 3} M^I
\epsilon_{abcde}v^{ab}v^{cd}{{\mathcal D}}_f v^{ef} \cr
 &~~+
{2\over 3} F^{Iab}\epsilon_{abcde}v^{cf} {{\mathcal D}}_f v^{de}
 +F^{Iab}\epsilon_{abcde}v^c_{~f}{{\mathcal D}}^d v^{ef}-{4 \over 3}F^{Iab}v_{ac}v^{cd}v_{db}\cr
 & ~~\left. -{1 \over 3} F^{Iab}
v_{ab}v^2 +4 M^I v_{ab} v^{bc}v_{cd}v^{da}-M^I (v^2)^2\right)~,
\label{eq:fourder}
\end{split}
\end{equation}
where the Weyl tensor is
\[
C_{abcd}=R_{abcd}-{2\over3}\left(g_{a[c}R_{d]b}-g_{b[c}R_{d]a}\right)+{1\over6}g_{a[c}g_{d]b}R~.
\]
The first term is the mixed gauge-gravity anomaly (\ref{eq:ano}). The second term includes
a more conventional {\it Riemann}-squared term, which was expected to be in the
same supermultiplet. The remaining terms are various matter terms and gravity-matter 
couplings. All of these are of the same order as the first two terms and so must be taken
properly into account. 

The complete four-derivative Lagrangian is clearly quite complicated and it would be
overwhelming to solve the corresponding equations of motion by brute force. However, 
we have already determined most features of the solution using general principles. 
All that is needed is a single equation of motion to find the metric factor $U$. 
The simplest is to use the equation of motion for the auxiliary $D$ field
\[
{\mathcal N} = 1 - {c_{2I}\over 72} \left( F^I_{ab} v^{ab} + M^I D \right)~.
\]
Inserting the expressions for the various fields in terms of charges and 
the metric factor we find
\begin{equation}
e^{-6U} = {1\over 6} c_{IJK} H^I H^J H^K + {c_{2I} \over 24} \left( \nabla H^I \nabla U + 2H^I \nabla^2 U\right)~.
\label{eq:defsg}
\end{equation}
The two derivative theory corresponds to keeping only the first term on the right hand side
and then the equation gives the metric factor $U$ explicitly. In the four derivative theory 
the equation has become a second order differential equation for the metric factor $U$. 
This describes the deformation
of the standard special geometry constraint (\ref{eq:specgeom})
due to higher derivative corrections. 

\subsection{Attractor Solution}
In the introduction we noted that, if the string solution is regular, the near
string geometry will be AdS$_3\times S^2$. The near string region is isolated by 
omitting the constant term in the harmonic functions
\[
H^I = p^I/2r~,
\]
and a AdS$_3\times S^2$ {\it ansatz} amounts to 
\[
e^{-6U} =  \ell^3_S/r^3~.
\]
We have for definiteness written the {\it ansatz} in terms of the $\ell_S$, the radius of the $S^2$, 
but supersymmetry implies $U_1=U_2$ which amounts to 
\[
\ell_A =2 \ell_S~,
\]
so we could just as well have used $\ell_A$, the radius of AdS$_3$. 

Now, inserting the {\it ansatz} into the deformed special geometry constraint (\ref{eq:defsg})
we see that the attractor {\it ansatz} is an exact solution with the identification 
of the $\ell_S$ as
\begin{equation}
\ell_S^3 = {1\over 8} \left( {1\over 6} c_{IJK} p^I p^J p^K + {1\over 12} c_{2I}p^I\right)~.
\label{eq:ls}
\end{equation}
The last term in this equation constitutes a correction to the scale $\ell_S$. In the
case of a string that is classically singular $c_{IJK} p^I p^J p^K=0$ and so this term
corrects the scale $\ell_S$ to a finite value. In this case the higher derivative terms 
have therefore resolved a singularity. 

\subsection{c-extremization}
In two-derivative gravity the central charge and the scale of the AdS$_3$ space
are related by the Brown-Henneaux formula (\ref{eq:dg}). Indeed this was 
a source of intuition that the singularity should be resolved. At this point have 
seen how derivative corrections modify the spacetime solution but we have yet 
to compute the corrected central charge. 

The total central charge $c={1\over 2}(c_R + c_L)$ is the trace anomaly of the  
two-dimensional boundary CFT. The holographic manifestation of the trace anomaly 
is an anomalous variation of the on-shell action under rescaling of the boundary 
metric. This variation is essentially the on-shell action itself so the general central
charge, including higher derivative corrections, equals the on-shell action of the 
attractor solution, up to a known overall numerical factor \cite{Kraus:2005vz}. 

Evaluating the full action (\ref{eq:lag0}), (\ref{eq:fourder}) on the attractor solution 
determined in the previous subsection we find 
\begin{equation}
c = - 12\ell^3_A\ell^2_S ({\mathcal L}_0 + {\mathcal L}_1) = c_{IJK} p^I p^J p^K + {3\over 4}c_{2I}p^I~,
\label{eq:cfct}
\end{equation}
in agreement with the exact central charges (\ref{eq:cd}) determined by anomaly inflow. 
More generally one can verify that the central charge (\ref{eq:cfct}) is extremized upon 
variation of the defining parameters of the attractor geometry. 
These agreements are sensitive consistency checks because the on-shell action 
depends on most terms in the actions (\ref{eq:lag0}), (\ref{eq:fourder}). 

\subsection{The interpolating solution}
We have focussed so far on the behavior of the string solution in the important
near string region. We now determine the complete radial evolution of the
metric factor $U$ in the special case of the solitonic
heterotic string, represented as $p$ $M5$-branes wrapping the $K3$ of 
$CY_3=K3\times T^2$.

To do this we need to solve the deformed special geometry constraint (\ref{eq:defsg}). 
This is a nonlinear second order differential equation which cannot generally
be solved analytically. However, we can proceed by solving one region at a time and
then patch together partial results to determine the complete solution: first 
consider the attractor solution applicable at 
$r\sim p^{1/3}$ and extend it perturbatively to larger distances; and then consider
the semiclassical solution around flat space, valid at $r\sim p$, and extend perturbatively
towards smaller distances. Alternatively, we can simply find the numerical solution to the full
differential equation. The results are presented in figure 1. It is seen that the expansion 
out from the attractor solution matches up quite well with the expansion in from asymptotically 
flat space, with the numerical solution shadowing both. 

The details of these procedures are given in \cite{Castro:2007hc} 
(and the review \cite{Castro:2008ne}). For the purpose of this talk the
main point is simply that all this can be done, providing some confidence
that the near horizon region attaches smoothly on to flat space in a rather conventional
fashion. 

\begin{figure}[tb]
\includegraphics[width=7cm,height=5cm]{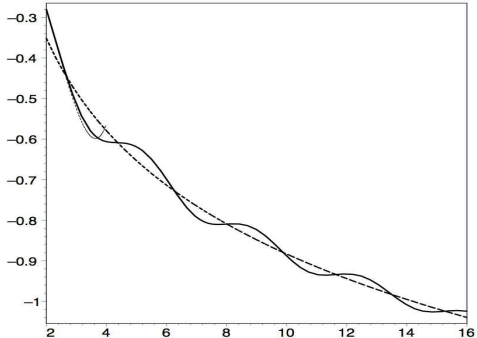}
\caption{Metric factor interpolating between near string region and asymptotic space. {\it Solid line}:
numerical solution. {\it Dashed line}: expansion around flat space. {\it Dotted line}: expansion 
around the near string solution.}
\label{firstfig}
\end{figure}

\subsection{Further corrections?}
We have included all corrections up to four derivative order in this work but one should ask
whether there might further corrections? As mentioned in the introduction, this question is
acute when corrections resolve a singularity for then the "corrections" are as important as 
the leading order terms. In other words, there is generally no systematic expansion parameter 
and so we must generally keep all orders in the Lagrangian. 

In our explicit contruction we have seen that the {\it form} of the solution is determined completely 
by supersymmetry. However, the final determination of the {\it scale} (\ref{eq:ls}) requires
solving the deformed special geometry constraint, an equation of motion which is expected to
receive further corrections. Thus our approach is {\it not} generally immune to further corrections.

Our result (\ref{eq:ls}) for the scale is an expansion with corrections controlled by an 
effective expansion parameter $1/p^2$. For large magnetic charge $p$, this structure 
precludes cancellation between different orders in the derivative expansion. 
However, for small black holes there is an additional challenge: some of the moduli are 
inherently small so higher order in inverse moduli can compensate for additional 
derivatives. For example, the dual heterotic string has strong coupling in the near 
string region, as one expects for a M5-brane. Therefore, 
it is not sufficient to consider solutions 
to the classical equations of motion as we do; quantum fluctuations around the saddle 
points may be significant. 

The issue of quantum fluctuations can be circumvented by dualizing the type IIA solitonic string 
to the heterotic frame. The metric factor near the string becomes
\[
e^{-6U} \to \ell^3/r^3~~~~,~ \ell= \sqrt{\alpha^\prime/2}~.
\]
The geometry of the heterotic string is thus AdS$_3\times S^2$, with the AdS and sphere
geometries of string 
scale. Note in particular that the scale of the geometry is independent  of the number 
of fundamental strings.
In this frame it is thus manifest that there is 
no systematic expansion parameter: the full geometry is stringy in nature. 
On the other hand, in the heterotic frame the dependence on the number of strings
enters exclusively through the coupling 
constant which is fixed as 
\[
g^{\rm het}_5 = 2^{-1/4} p^{-1/2}~,
\label{eq:ghet}
\]
so quantum corrections are under good control for a large number of strings. 

In the string geometry the precise definition of scale is ambiguous and one may try to 
maintain that central charge is a good proxy for scale. In that sense the higher derivative
corrections resolve the gravitational singularity, but with other notions of length
the resolution is merely qualitative, which are subject to further corrections of the same 
order as those that have been included.

\section{The holographic dual of the heterotic string}

The supergravity representation of the heterotic string is of string scale, but it is still reasonable 
to ask what the holographic 
dual is. The standard line of reasoning suggests that it should be a it should be a 
$D=1+1$ CFT with $(8,0)$ supersymmetry and R-symmetry at least SU(2). 
The classical super-isometry group of the solution is $OSp(*4|4)$ \cite{Lapan:2007jx},
so one might in fact
expect that the dual theory represents an affine extension of that group. The challenge is that
there exist no CFT with all these properties. There is simply too much supersymmetry. 

A possible resolution to this conundrum is suggested by the existence of nonlinear superconformal 
algebras (NSCA's) with the correct symmetries \cite{Henneaux:1999ib}. The nonlinearity of these 
algebras refer to the supercurrent OPE which takes the form
\[
G^i(z) G^j(0)  \sim {k\over\chi}{\delta^{ij}\over z^3} + 
{1\over\chi} {J^a ( T^a)^{ij}\over z^2} + {2T\delta^{ij}\over z}
+{1\over 2\chi}
{\partial_z J^a_z T^a\over z} + {1\over 2\chi k} {J^a_z J^b_z P^{ij}_{ab}\over z}~.
\]
The last term is bilinear in the currents and so represents an unfamiliar non-linearity. 
The nonlinear term depends on the tensor
\[
P^{ij}_{ab} = {1\over 2}\{ T^a, T^b\}^{ij} - 2\chi\delta_{ab}\delta^{ij}~,
\]
where $\chi$ is a constant depending on the group. This tensor is such that the 
non-linearities vanish for small groups like $SU(2)$ but not for larger $R$-symmetry 
groups. 
The nonlinear superconformal algebras \cite{Knizhnik:1986wc,Bershadsky:1986ms} 
are powerful but unfamiliar relatives to W-algebras.
In the current setting we consider multi-string states so the suggestion is that NSCA's are
important in string field theory. 

An intriguing feature of the NSCA's is that they give nonlinear formulae for the central charge. 
For example, the conformal extension of $OSp(*4|4)$ has central charge
\begin{equation}
c = - 12p( 1 + {3\over 2p})~.
\label{eq:ncact}
\end{equation}
The classical (large $p$) limit gives the Brown-Henneaux formula (except for the sign). 
However, $p\sim 1/g^2$ so the nonlinear algebra appears to identify nontrivial quantum 
corrections to the spacetime central charge. Interesting as this prospect may seem it 
cannot be emphasized enough that there are severe obstacles to this optimistic 
interpretation. The biggest problem is that the relevant algebras apparently
have no unitary representations. This problem is manifest in the expression (\ref{eq:ncact}) 
for the central charge, which is in fact negative. Unitarity violation is not only unacceptable on 
general grounds, it is also the wrong physics for a setting expected to be
extremely stable because of the high degree of supersymmetry. 

It is not clear what conclusion one should draw from this. On the one hand, the appearance
of NCSAs is tantalizing, and almost inevitable given the symmetries of the situation. 
On the other hand, the most straightforward implementation clearly needs to be modified.
This situation makes the challenge of finding a satisfying description
of the theory even more interesting. 

\section{Discussion}
In this talk we discussed the resolution of gravitational singularities with emphasis
on the case of 5D string solutions with AdS$_3\times S^2$ near string geometry. 
However, the techniques apply in many other cases where one seeks to resolve 
singularities, or to correct solutions that are already regular in the two-derivative theory. 
Some examples are:
\begin{itemize}
\item
{\it Black holes in five dimensions} with AdS$_2\times S^3$ near horizon geometry. The
solutions are supported by electric charges so the Bianchi identity is trivial. Instead one must
work out Gauss' law from the explicit action. Fortunately, it turns out that the result again
reduces to a harmonic equation and the complete solution 
can be determined explicitly.
\item
{\it Rotating black holes} in five dimensions have less symmetry and are correspondingly
more complicated. Although all terms in the four-derivative action (\ref{eq:fourder}) now
contribute, Gauss'
law remains integrable, and again the complete solution can be determined. 
\item
{\it Rotating black holes on Taub-NUT base space} are more involved yet but, remarkably,
all equations remain integrable in much the same way as the simpler case. The significance
of Taub-NUT is that its "cigar"-form interpolates between a central $R^4$ region and asymptotic
$R^3\times S^1$. Thus these 5D solutions on Taub-NUT include the 4D black holes
that have been considered previously. The detailed comparison reveals discrepancies
which can be traced to subtleties in the definition of charge in the presence of higher derivative
terms.  
\end{itemize} 
These results and others were worked out in \cite{Castro:2007sd,Castro:2007hc,Castro:2007ci}
and were recently reviewed in detail in \cite{Castro:2008ne}. The combined body of examples 
verify previous indirect arguments by explicit computation, sort out discrepancies in those 
arguments, and uncovers numerous new results, including many still await interpretation. 
The explicit construction of supersymmetric solutions to five dimensional theories 
with higher derivatives has thus proven a fruitful endeavor.
 
Among many open problems, the most far-reaching involve the generalizations
to other dimensions, including 10 and 11 dimensions, and theories with more general 
matter content. In such more general contexts there are often very different motivations for
considering higher derivative corrections like, for example, the recent reports of the possible
finiteness of $N=8$ supergravity\cite{Green:2007zzb,Bern:2007xj}. Many of the higher 
derivative terms identified for different purposes could be employed also in the context of the 
resolution of gravitational singularities. It is therefore reasonable to expect a fruitful 
interplay between different perspectives on higher derivative corrections to gravitational
theories.

\section*{Acknowledgments}
I am grateful to A. Castro, J. Davis, P. Kraus, and A. Shah for collaboration on the work
reported on here \cite{Castro:2007sd,Castro:2007hc,Castro:2007ci,Kraus:2007vu,Castro:2008ne}.
I also thank A. Giveon and K. Hanaki for discussions. 
This work was supported by DOE under grant DE-FG02-95ER40899.

\bibliographystyle{amsalpha}

\end{document}